\RequirePackage{ifpdf}
\ifpdf 
\documentclass[pdftex]{sigma}
\else
\documentclass{sigma}
\fi

\begin{document}

\allowdisplaybreaks

\renewcommand{\PaperNumber}{045}

\FirstPageHeading

\renewcommand{\thefootnote}{$\star$}

\ShortArticleName{Spectral Curves of Operators with Elliptic
Coef\/f\/icients}

\ArticleName{Spectral Curves of Operators\\ with Elliptic
Coef\/f\/icients\footnote{This paper is a contribution to the
Vadim Kuznetsov Memorial Issue `Integrable Systems and Related
Topics'. The full collection is available at
\href{http://www.emis.de/journals/SIGMA/kuznetsov.html}{http://www.emis.de/journals/SIGMA/kuznetsov.html}}}

\renewcommand{\thefootnote}{\arabic{footnote}}
\setcounter{footnote}{0}

\Author{J. Chris EILBECK~$^\dag$, Victor Z. ENOLSKI~$^\ddag$\footnote{Wishes to express thanks
to the  MISGAM project for support of his research
visit to SISSA (Trieste) in 2006.} and
Emma PREVIATO~$^\S$}

\AuthorNameForHeading{J.C. Eilbeck, V.Z. Enolski and E. Previato}

\Address{$^\dag$~The Maxwell Institute and Department of
Mathematics,
Heriot-Watt University,\\
$\phantom{^\dag}$~Edinburgh,    UK EH14 4AS}
\EmailD{\href{mailto:J.C.Eilbeck@hw.ac.uk}{J.C.Eilbeck@hw.ac.uk}}

\Address{$^\ddag$~Institute of Magnetism, 36 Vernadski Str.,
Kyiv-142, Ukraine}

\EmailD{\href{mailto:vze@ma.hw.ac.uk}{vze@ma.hw.ac.uk}}

\Address{$^\S$~Department of Mathematics and Statistics, Boston
University, Boston MA 02215-2411, USA}
\EmailD{\href{mailto:ep@bu.edu}{ep@bu.edu}}

\ArticleDates{Received November 21, 2006, in f\/inal form February
16, 2007; Published online March 12, 2007}

\Abstract{A computer-algebra aided method is carried out, for
determining geometric objects associated to dif\/ferential
operators that satisfy the elliptic ansatz. This results in
examples of Lam\'e curves with double reduction and in the
explicit reduction of the theta function of a Halphen curve.}


\Classification{33E05; 34L10; 14H42; 14H45}

\begin{flushright}
\textit{Dedicated to the memory of Vadim Kuznetsov}
\end{flushright}

\section{Introduction}\label{sec1}

The classical theory of reduction, initiated by Weierstrass, has
found modern applications to the f\/ields of Integrable Systems
and Number Theory, to name but two. In this short note we only
address specif\/ic cases, providing minimal historical references,
listing the steps that we devised, and exhibiting some new
explicit formulas. A full-length discussion of original
motivation, theoretical advancements and modern applications would
take more than one book to present fairly, and again, we choose to
provide one (two-part) reference only, which is recent and
captures our point of view \cite{[BE1],[BE2]}.

Our point of departure is rooted in the classical theory of
Ordinary Dif\/ferential Equations (ODEs).  At a time when activity
in the study of elliptic functions was most intense,  Halphen,
Hermite and Lam\'e (among many others) obtained deep results in
the spectral theory of linear dif\/ferential operators with
elliptic coef\/f\/icients. Using dif\/ferential algebra, Burchnall
and Chaundy described the spectrum (by which we mean the joint
spectrum of the commuting operators) of those operators that are
now called algebro-geometric, and some non-linear Partial
Dif\/ferential Equations (PDEs) satisf\/ied by their
coef\/f\/icients under isospectral deformations along `time'
f\/lows, $t_1,\ldots ,t_g$, where $g$ is the genus of the spectral
ring. We seek algorithms that, starting with an Ordinary
Dif\/ferential Operator (ODO) with elliptic coef\/f\/icients,
produce an algebraic equation for an af\/f\/ine plane model of the
spectral curve (Section \ref{sec2}). This problem turns out to
merge with Weierstrass' question \cite{[Ke]}, Appendix C:

{\it Give a condition that an algebraic function $f(x)$ must
satisfy if among the integrals
\begin{gather*}
\int F(x,f(x))dx,
\end{gather*}
where $F$ is a rational function of $x$ and $f(x)$, there exist
some that can be transformed into elliptic integrals}.

\looseness=1 This means that the curve $X$ whose function f\/ield
is ${\Bbb C} (x,f(x))$ has a holomorphic dif\/ferential which is
elliptic, or equivalently, Jac\,$X$ contains an elliptic curve.
Weierstrass gave a condition in terms of the period matrix of $X$.
Burchnall--Chaundy spectral curves of operators with elliptic
coef\/f\/icients automatically come with this reduction property,
because the dif\/ferential-algebra theory guarantees that all the
coef\/f\/icients of the spectral ring are doubly-periodic with
respect to the same lattice; the PDEs mentioned above have
solutions which are called `elliptic solitons'.  A more special
condition is that there be a `multiple' reduction, namely another
periodicity with respect to a (possibly non-isomorphic) lattice,
in which case a curve of genus three has Jacobian isogenous to the
product of three elliptic curves, and this is what we found in
\cite{[EEP]}. Another dif\/f\/icult question is whether the
Jacobian is actually complex-isomorphic (without the principal
polarization, since Jacobians are indecomposable) to the product
of three elliptic curves. This is the case for the Klein curve
\cite{[B]}; we do not know if it is for the Halphen curve in
\cite{[EEP]}. In genus two, the situation is only apparently
simpler: the question of complex-isomorphism to a product has only
recently been settled \cite{[E]}, and that, in an analytic but not
in an algebraic language; the question of loci of genus-2 curves
which cover ($n:1$) an elliptic curve remains largely open
(despite recent progress, much of it based on computer
algeb\-ra~\cite{[SV]}).

\looseness=1 In Section \ref{sec3} of this paper, we provide
progress towards such questions for the case of  Lam\'e curves,
which are hyperelliptic: the curves are known, at least for small
genus, but we report a double reduction (equianharmonic case) that
we have not seen elsewhere; in Section~\ref{sec4}, we treat the
genus~3 non-hyperelliptic case. Our contribution to a method
consists of two parts. First, we  provide algebraic maps between
the higher-genus and the genus-one curves, by giving a search
algorithm that identif\/ies genus-1 integrals,  the very question
posed by Weierstrass. Se\-cond, using classical reduction theory
as revisited by Martens~\cite{[Ma]}, we write the theta function
for a case of reduction in terms of genus-1 theta functions (or
the associate sigma functions). We connect the two issues, by
calculating explicitly the period matrix for some curve that
admits reduction, given algebraically.  It is a rare event, that a
period matrix can be calculated from an algebraic equation, since
the two in general are transcendental functions of each other. The
method consists in the traditional one \cite{[B],[Q],[S]} of
choosing a suitable homology basis and calculating the covering
action on it. We believe our result on the period matrix of the
Halphen curve and its reduction to be new. We also address the
question of ef\/fectivization of the solution of the KP equation,
or hierarchy; for this to be complete, however, we need not only
to reduce the theta function, but also the integrals of
dif\/ferentials of the second~kind.

We dedicate this small note to the memory of Vadim Kuznetsov. It
was one of Vadim's many projects to f\/ind a systematic answer to
the problem of giving a canonical form for dif\/ferential
equations that have a basis of solutions
 polynomial in elliptic functions, generalizing
the theory of elliptical harmonics given in the last chapter of
\cite{[WW]}, and which we have implemented in Section \ref{sec3},
to study the Lam\'e curves.  In fact, one among us (VZE)
co-authored one of the last papers of Vadim \cite{[YAKE]}, devoted
to a problem that arises in physics and was f\/irst identif\/ied
by him; some of the ideas that we describe below were to be used
in the further development of that subject. The loss of Vadim as
friend and mathematician has af\/fected us deeply. Personal
memories and testimonials are collected in other parts of this
issue; here we   only say that it was our great privilege to  know
him.

\section{Elliptic covers}\label{sec2}

Curves that admit a f\/inite-to-one map to an elliptic curve
(`Elliptic Covers') are special, and are a topic of extensive
study in algebraic geometry and number theory. They have come to
the fore in the theory of integrable equations, in particular
their algebro-geometric solutions, for the reason that a yet more
special class of such curves provides solutions that are
doubly-periodic in one variable (`Elliptic Solitons').

Our point of view on the latter class of curves is based on the
Burchnall--Chaundy theory,  {\it via} which, we connect the
problem to the  theory of ODOs with elliptic coef\/f\/icients,
much studied in the nineteenth century (Lam\'e, Hermite, Halphen,
e.g.). It is only when such an operator has a large commutator that
this theory connects with that of integrable equations. While the
nineteenth-century point of view could not have phrased the
special property in this way, the related question that was
addressed was, when does the series expansion of solutions, in
some suitable local coordinate on the curve (cf.~Remark 1 in
Subsection~3.1)
 terminate?
The early work turned up important, but somehow {\it ad hoc},
properties that characterize algebro-geometric operators.

\subsection{Challenge I}
Detect (larger) classes of algebro-geometric operators.

We do not address this issue here.  A CAS-based project would be
to test whether the operator
\begin{gather*}
L=\partial^3 -\dfrac{4}{3}n^2\wp\partial +
\dfrac{2n(n-3)(4n+3)}{27}\wp'
\end{gather*}
has centralizer larger than ${\Bbb C}[L]$: we ask the question
since Halphen \cite{[H]} noticed that the equation $L=0$ has
solutions that can be expressed in terms of elliptic functions. A
more far-reaching project would construct
elliptic-coef\/f\/icient, f\/inite-gap operators starting with
curves that were shown to be elliptic covers, choosing a local
coordinate whose suitable power is a function pulled back from the
elliptic curve, to give rise to the commutative ring of ODOs,
according to the Burchnall--Chaundy--Krichever inverse spectral
theory. Curves of genus 3 that are known to be elliptic $3:1$ and $2:1$ covers, respectively, are Klein's and
Kowalevski's quartic curves:
\begin{gather*}
zx^3+xy^3+yz^3=0,
\\[1ex]
(z^2-f_2)^2=4xy(ax+by)(cx+dy)
\end{gather*}
in projective coordinates, with $f_2$ a form of degree 2 in $x$
and $y$ (such that the quartic is non-singular). Notably, Klein's
curve (the only curve of genus 3 that has maximum number of
automorphisms, 168) has Jacobian which is isomorphic to the
product of three (isomorphic) elliptic curves \cite{[B], [Pr],
[RL]}; Kowalevski, as part of her thesis, classified the (non-hyperelliptic) curves of genus 3 that cover
$2:1$ an elliptic curve (Klein's curve being a special case).

\subsection{Challenge II}
Find an equation for the spectral curve.

This can be found once an algebro-geometric ODO is given.  While
the Burchnall--Chaundy theory allows one in principle to write the
equation of the curve as a dif\/ferential resultant~\cite{[Pre]},
this would be unwieldy for all but the simplest cases: f\/irst,
one would have to solve the ODEs of commutation $[L,B]=0$  for an
unknown operator $B$ of order co-prime with $L$ (the simplest case
of a two-generator centralizer); then, once $B$ is known,
calculate the determinant of the $(m+n)\times (m+n)$ matrix that
detects the common eigenfunctions of $L-\lambda$ and $B-\mu$
\cite{[EE2]}.

Instead, it is much more ef\/f\/icient to adapt Hermite's (or
Halphen's) ansatz; in the case of Lam\'e's equation, the case of
solutions `written in f\/inite form' is treated in \cite{[WW]};
more generally, assume that an eigenfunction can be written in
terms of:
\begin{gather*}
\Psi (x;\alpha )=e^{kx} \sum_{j=0}^{n-1} a_j(x,\mu, k)
\dfrac{d^j}{dx^j}\Phi (x;\alpha )
\end{gather*}
solving  a given spectral problem: $L\Psi (x)=z\Psi (x)$, where:
\begin{gather*}
\Phi (x,\alpha )=-\dfrac{\sigma (x-\alpha)}{\sigma (\alpha)\sigma
(x)}e^{\zeta (\alpha )x}
\end{gather*}
and $\alpha$ is a complex number viewed as a point of the elliptic
curve $\nu^2=4\mu^3-g_2\mu -g_3$.

By expanding  $\wp (x)$ and  $\phi (x,\alpha )$ near $x=0$ and
comparing coef\/f\/icients, the $a_j$ can be written in terms of
($k,z$) and taking the resultant of the compatibility conditions
gives an equation of the spectral  curve in the ($k,z$)-plane
\cite{[EE1]}.

\subsection{Challenge III}
Express the eigenfunctions in terms of the theta function of the
curve.

\subsection{Challenge IV}
Turn on the KP (isospectral-)time deformations $t_i$ and express
the time-dependence of the coef\/f\/icients of the operators (in
particular, one such coef\/f\/icient is an exact solution of the
KP equation).  The expression of a KP solution in terms of
elliptic functions is part of the programme sometimes referred to
as `ef\/fectivization'.

Our approach to such `challenges' is two-fold: on one hand we use
the explicit expression of the higher-genus sigma functions that
were classically proposed by Klein and Baker, but revived,
substantively generalized, and brought into usable form at f\/irst
by one of the authors with collaborators (cf. e.g. \cite{[BEL]}).
On the other hand, we design computer algebra routines which allow
us to read, for example, coef\/f\/icients of multi-variable Taylor
expansions far enough to obtain the equations for the curves and
the KP solutions for small genus.

In this paper we present classes of examples, originating with classical
ansatzes of Lam\'e, Hermite and Halphen (specif\/ically
applied to elliptic solitons by Krichever \cite{[Kr]}) as well as
some detail of our general strategy.

\section{Hyperelliptic case}\label{sec3}

The Lam\'e equation has been vastly applied and vastly studied: we
refer to \cite{[M]}
for general information, and to the classical treatise \cite{[WW]}
for calculations that we shall need (for richer sources please
consult \cite{[GW1]}, and \cite{[GW2]} for dif\/ferent
perspectives). Our point of view is that the potentials
 $n(n+1)\wp (\xi )$ are f\/inite-gap when $n$ is an integer
(an adaptation of Ince's theorem to the complex-valued case) where
the equation is written in Weierstrass form:
\begin{gather}
\dfrac{d^{2}w(\xi)}{d\xi^2}-n (n+1)\wp(\xi)w(\xi)=zw(\xi)
\label{eq1}
\end{gather}
(for the rational form see \cite{[M]} or \cite{[WW]}).

\subsection{The spectral curve}
The Lam\'e curves are hyperelliptic since one of the commuting
dif\/ferential operators has order~2, and will have genus $n$ for
the operator in \eqref{eq1}. There is no `closed form' for general
$n$, but the equations have been found by several authors and
methods, cf.\ specif\/ically \cite{[BE1], [BE2]} and references
therein (work by Eilbeck {\it et al.}, by Enol'skii and
N.A.~Kostov, by A.~Treibich and by J.-L.~Verdier is there
referenced), \cite{[GV], [M], [MS], [T4]}.  In particular,
\cite{[M]} adopts a method of inserting ansatzes in the equation
and comparing expansions similar to ours, and tabulates the
equations for $1\le g\le 8$ (in principle, a recursive calculation
will produce them for any genus), so we do not present our table
($1\le g\le 10$), but use $g=3$ to exemplify  our treatment of
reduction in Subsection~3.2 below.

\begin{remark}
To motivate the project that Vadim Kuznetsov had wanted to design,
as recalled in the introduction, we mention in sketch the steps of
our strategy to f\/ind the spectral curves. The key ansatz is the
f\/inite expansion of the eigenfunctions; we used a formula from
\cite{[WW]}  in the case of odd $n$ and the (equivalent) table in
\cite{[A]} for even $n$ just for bookkeeping purposes;
respectively,
\begin{gather*}
w(\xi) =\sqrt {\wp(\xi)-e_{{1}}}\sqrt {\wp(\xi)-e_{2}}
\sum_{r=0}^{(n-3)/2} b_{{r}}(\wp(\xi ) -e_{3})^{n/2-r-1}, \qquad
w(\xi) = \sum_{r=0}^{n/2} A_{{r}}(-\wp(\xi ))^{r}.
\end{gather*}
It is then a matter of f\/inding the coef\/f\/icients $b_r, \
A_r$. For this, we insert $w$ in (1) and expand in powers of $\wp
(\xi )$ or $(\wp (\xi ) - e_i)^{1/2}$, depending on the various
cases; we
 also substitute for the $e_i$ in terms
of the modular functions $g_2$ and $g_3$. By comparing
coef\/f\/icients in the expansions we obtain a set of equations
for the $b_r$ or $A_r$, and we solve the compatibility condition
of these equations. The Lam\'e curve $
f_s(z)\prod\limits_{i=1}^3f_i(z) $ is the product of 4 factors,
one $f_s(z)$ symmetric in the~$e_i$, i.e.\ just depending on $g_2$
and $g_3$, and three $f_i(z)$ which depend on one of the $e_i$ and
$g_2$ and $g_3$, which we found in turn. At the end, to be sure,
for a given root $z$ of one of these factors, we can determine the
$b_r$, resp. $A_r$
 up to a common factor, and write the eigenfunction
as a  polynomial in $\wp(\xi)$. The f\/initeness condition we used
in expanding the eigenfunctions (a sort of Halphen's ansatz, cf.
Section 1) characterizes f\/inite-gap operators, in
Burchnall--Chaundy terms, the ones that have large centralizers.
It is worth quoting a speculation \cite{[SW]} p. 34 which regards
f\/initeness in the local parameter $z^{-1}$: `We do not know an
altogether satisfying description of the desired class ${\cal
C}^{(n)}$; roughly speaking, it consists of the operators whose
formal Baker functions converge for large $z$.'
\end{remark}

\subsection{Cover equations and reduction}
  Even though the Lam\'e curves are
elliptic covers {\it a priori}, it is not easy to f\/ind the
elliptic reduction, which will correspond to a holomorphic
dif\/ferential of smallest order at the point $\infty$ in
$(w,z)$-coordinates\footnote{A dif\/ferent and very ef\/f\/icient
method was developed  by Takemura in the beautiful series
\cite{[T1], [T2], [T3], [T4]} where elliptic-hyperelliptic
formulas are found by comparing the monodromies on the two
curves.};
 by running through a linear-algebra
elimination for a basis of holomorphic dif\/ferentials, expressed
in terms of rational functions of $(w,z)$ (here again expansions
provide the initial guess, lest the programs exhaust computational
power), we exhibited the elliptic dif\/ferential for all our
curves. Note that this gives the degree of the cover, not known
{\it a priori}. It is worth pointing out
 that more than $n$ covers might be found (for genus $n$):
indeed, according to a classical result (Bolza, Poincar\'e, cf.
\cite{[K]} where a modern proof is given), if an abelian surface
contains more than one elliptic subgroup, then it contains
inf\/initely many. Note, however, that it can only contain a
f\/inite number for any given degree (the intersection number with
a principal polarization) because the N\'eron--Severi group is
f\/initely generated \cite{[K]}. This degree translates into the
degree of the elliptic cover.  We are not claiming to have found
all of them, or that those which we found be minimal. For example,
for genus $n=2$ (where the reduction is known since the nineteenth
century), f\/inding $f_s=z^2-3g_2$, $f_i=z+3e_i$, in expanded form
the curve is:
\begin{gather*}
{w}^{2}=- (3g_{2}-{z}^{2})(27g_{3}-4{z}^{3}+9zg_{2}).
\end{gather*}
The holomorphic dif\/ferential is
\begin{gather*}
\dfrac{d\wp}{\wp'}=3z\dfrac{dz}{w},
\end{gather*}
and a second reduction (predicted by  a classical theorem of
Picard that splits any abelian subvariety up to isogeny) is given
by the algebraic map to:
\begin{gather*}
  {\wp'}^2=4{\wp}^3-G_2\wp-G_3,
  \qquad
  G_2=\frac{27}{4}(g_2 ^3+9g_3 ^2),\qquad G_3= \frac{243}{8} g_3 (3g_3 ^2- g_2 ^3),
  \\
  \wp= -\frac14\,(4z^3-9g_2 z-9g_3),
  \qquad
  \wp'= -\frac12\,w\,(4z^2-3g_2).
\end{gather*}
\looseness=1 This map induces the following reduction of the
holomorphic dif\/ferential to an elliptic dif\/fe\-rential:
\begin{gather*}
\frac{d\wp}{\wp'}=-3\frac{d}{w}.
\end{gather*}

Notice then that in the equianharmonic case $g_2=0$ the curve is
singular, $w^2=-(4z^3-27g_3)z^2$ and we have the following
birational map to the elliptic curve:
\begin{gather*}
{\wp'}^2=4{\wp}^3-G_3,\qquad \wp  =  -\frac{z}{4}, \qquad \wp'=
\frac{-2w}{4z},\qquad
 G_3 = -\frac{27}{64} g_3
\end{gather*}
and dif\/ferential $d\wp/\wp' = zdz/2w.$ This motivated us to
investigate the equianharmonic case further.

What we believe to be new is the following observation.  We found
it surprising that there always be at least a second reduction in
the case the elliptic curve is equianharmonic (the unique one that
has an automorphism of order three), but see no theoretical reason
to expect that. If this were to hold for all $n$, one could
further ask whether these curves support KdV solutions elliptic
both in the f\/irst and second time variables (a private
communication by A. Veselov gives such an indication, to the best
of our understanding). The issue of periodicity in the second KdV
variable (`time')  was put forth in \cite{[Sm]}, and recently
reprised in \cite{[FT]}, but the double periodicity does not seem
to have been addressed so  far. We exhibit three covers in the
$g=3$ case (N.B.~The third one is not, in general,  over the
equianharmonic curve); as well, we found two covers in the
equianharmonic, $n=4,5$ cases, but we only provide a summary table
which suggests the general pattern for two covers in the
equianharmonic case.

For {$n=3$}, $f_s=z$,   $f_i=z^2 - 6z e_i + 45 e_i^2 - 15g_2$,
 the curve is
\begin{gather*}
{w}^{2}= \big(2376\,{z}^{3}g_{3}-36450\,zg_{2}g_{3}+504\,g_{{2
}}{z}^{4}-91125\,{g_{3}}^{2}-16\,{z}^{6}+3375\,{g_{2}}^{3}-4185\,{
g_{2}}^{2}{z}^{2}\big) z,
\end{gather*}
the f\/irst cover is given by the algebraic map
\begin{gather*}
 \wp=-\frac{ (-84375 g_2^3+2278125 g_3^2+303750 g_3 g_2
    z-3375 g_2^2 z^2-27000 z^3 g_3+360 z^4 g_2+16 z^6)}{36\,z\, (-75 g_2+4 z^2)^2 },
\\[1ex]
  \wp'= \frac{w\, (16 z^6-1800 z^4 g_2-16875 g_2^2
    z^2+421875 g_2^3+54000 z^3 g_3-11390625 g_3^2)}{108\,z^2\, (-75 g_2+4z^2)^3}.
\end{gather*}
This map induces the following reduction of the holomorphic
dif\/ferential to an elliptic dif\/fe\-rential:
\begin{gather*}
\frac{d\wp}{\wp'}=-3(4 z^2-15 g_2)\frac{dz}{w}.
\end{gather*}

In the equianharmonic case the above reduces to the following
cover of the curve ${\wp'}^2=4{\wp}^3-g_3$:
\begin{gather*}
\wp=-\frac{1}{576}\frac{(16 z^6-27000 z^3 g_3+2278125
g_3^2)}{z^5},
\\[1ex]
\wp'=\frac{1}{6912} \frac{w(16 z^6+54000 z^3
g_3-11390625g_3^2)}{z^8}.
\end{gather*}
This map induces the following reduction of the holomorphic
dif\/ferential to an elliptic dif\/fe\-rential:
\begin{gather*}
\frac{d\wp}{\wp'}=-12\frac{z^2dz}{w}.
\end{gather*}

Continuing in the equianharmonic case, the second cover is given
by the following algebraic map to:
\begin{gather*}
{\wp'}^2=4{\wp}^3-G_3,\qquad
  G_3=\frac{387420489}{4} g_3^5,
\\[1ex]
 \wp=-\frac{1}{6} \frac{3645 g_3^2-1080 z^3 g_3+16 z^6}{z},
\qquad \wp'= \frac{1}{32} \frac{w(-729 g_3^2-432 z^3
g_3+16z^6)}{z^2}.
\end{gather*}
This map induces the following reduction of the holomorphic
dif\/ferential to an elliptic dif\/fe\-rential:
\begin{gather*}
\frac{d\wp}{\wp'}=-10\frac{dz}{w}.
\end{gather*}

The third cover is given by the algebraic map  to the
(non-equianharmonic) curve
\begin{gather*}
{\wp'}^2=4{\wp}^3-G_2\wp -G_3\equiv 4\,(\wp-198
g_3)\,(12717g_3^2+198\wp g_3+\wp^2),
\\[1ex]
 G_2= 105948 g_3^2,\qquad G_3 = 10071864 g_3^3,
\qquad \wp=198 g_3-4 z^3,\qquad
  \wp'= 4 w z.
\end{gather*}
This map induces the following reduction of the holomorphic
dif\/ferential to an elliptic dif\/fe\-rential:
\begin{gather*}
\frac{d\wp}{\wp'}=-3\frac{z{d}z}{w}.
\end{gather*}

For the curves:
\begin{center}\renewcommand{\arraystretch}{1.02}
\begin{tabular}{cll}
\hline \multicolumn{1}{c|}{\parbox{8mm}{\centerline{$n$}}}&
\multicolumn{1}{c|}{$f_s$} & \multicolumn{1}{c}{$f_i$}
\\
\hline 1& $ f_s=1$ & $f_i=z-e_i$
\\
2& $f_s=z^2-3g_2$&$f_i=z+3e_i$
\\
3& $f_s=z$&    $f_i=z^2 - 6z e_i + 45 e_i^2 - 15g_2$
\\
4&  $f_s=z^3 - 52 g_2 z + 560g_3$& $f_i=z^2+10 z e_i-7g_2-
35e_i^2$
\\
5& $f_s=z^2-27g_2$& $f_i={z^3-15 z^2e_2 + (315
e_2^2-132g_2)z+675e_2^3+540g_3}$
\\
\hline
\end{tabular}
\end{center}
we summarize the equianharmonic covers in the table below (note:
there is some small numerical discrepancy, e.g., for $n=2$ our
choice of normalization corresponds to the $-2w$ of this table,
but we f\/ind it more useful to have the normalization below for
recognizing a pattern). The pairs appearing in the second column
are the orders of the commuting dif\/ferential operators.

\begin{center}\renewcommand{\arraystretch}{1.9}
\begin{tabular}{cllllll}
\hline \multicolumn{1}{c|}{\parbox{8mm}{\centerline{$n$}}}&
\multicolumn{1}{c|}{$(-,-)$} & \multicolumn{1}{c|}{$\wp$} &
\multicolumn{1}{c|}{$\wp'$} &
\multicolumn{1}{c|}{$\dfrac{d\wp}{\wp'}$\bsep{1.7ex}} &
\multicolumn{1}{c|}{$G_2$} & \multicolumn{1}{c}{$G_3$}
\\
\hline 2& (2,5) &$9g_3-4z^3$ &$ 4wz^2$ &$-3\dfrac{{d}z}{w}$
&$486g_3^2$ &$ 729g_3^3$
\\
3 &(2,7) &$198 g_3-4 z^3$ &$4wz$ &$-3z\dfrac{{d}z}{w}$
&$10594g_3^2$ &$10071864 g_3^3$
\\
4 &(2,9) &$1430 g_3-4 z^3$ &$4w$ &$-3z^2\dfrac{{d}z}{w}$
&$6342300g_3^2$ &$3011499000g_3^3$
\\
5 &(2,11) &$6435 g_3-4 z^3$ &$4\dfrac{w}z$ &$-3z^3
\dfrac{{d}z}{w}$ &$28674000g_3^2$ &$ 14791410000
g_3^3$\bsep{1.5ex}
\\
\hline
\end{tabular}
\end{center}
In factorized form the coef\/f\/icients are
\begin{center}\renewcommand{\arraystretch}{1.9}
\begin{tabular}{llllll}
\hline \multicolumn{1}{c|}{$(-,-)$} & \multicolumn{1}{c|}{$\wp$} &
\multicolumn{1}{c|}{$\wp'$} &
\multicolumn{1}{c|}{$\dfrac{d\wp}{\wp'}$\bsep{1.7ex}} &
\multicolumn{1}{c|}{$G_2$} & \multicolumn{1}{c}{$G_3$}
\\
\hline (2,5) &$3^2g_3-4z^3$ &$4wz^2$ &$-3\dfrac{dz}{w}$ &$2\cdot
3^5 g_3^2$ &$3^6 g_3^3$
\\
(2,7) &$2\cdot 3^2\cdot 11 g_3-4 z^3$ &$4wz$ &$-3z\dfrac{dz}{w}$
&$2^2 \cdot 3^5\cdot 109g_3^2$ &$2^3\cdot 3^6\cdot 11\cdot
157g_3^3$
\\
(2,9) &$2\cdot 5\cdot 11\cdot 13g_3-4 z^3$ &$4w$
&$-3z^2\dfrac{dz}{w}$ &$2^2\cdot3^7\cdot5^2\cdot29g_3^2$
&$2^3\cdot3^{11}\cdot5^3\cdot 17g_3^3$
\\
(2,11) &$3^2\cdot 5\cdot 11\cdot 13g_3-4z^3$ &$4\dfrac{w}{z}$
&$-3z^3\dfrac{dz}w$ &$2^4\cdot 3^5\cdot 5^3\cdot 59g_3^2$
&$2^4\cdot 3^6\cdot5^4\cdot 2029g_3^3$\bsep{1.7ex}
\\
\hline
\end{tabular}
\end{center}
The hyperelliptic curves are singular for $g=2,5,8$; this is
important because it makes it possible to explicitly calculate the
motion of the poles of the KdV solutions
(Calogero--Moser--Krichever system). When all the KdV or KP
dif\/ferentials (of increasing order of pole) are reducible, of
which case we have examples in genus 3, both hyperelliptic as we
have seen, and non (Section \ref{sec4}),
 the Calogero--Moser--Krichever system is periodic in all
time directions; we believe this situation had not been previously
detected.

\subsection{The theta function}
The next challenge we discuss is the
 calculation of  the coef\/f\/icients of the ODOs
in terms of theta functions; while we do not present a result that
could not have been derived by classical methods, we streamlined a
two-step  procedure: f\/irstly, we use Martens' thorough
calculation of the action of the symplectic group on a reducible
period matrix \cite{[Ma]}, to write theta as a sum, then we use
transformation rules for the action under the symplectic group.
The step which is not obvious however precedes all this and is the
calculation of the explicit period matrix based on a reduced basis
of holomorphic dif\/ferentials. We give the result for one
specif\/ic curve as an example. For  the curve of genus 2:
\begin{gather*}
w^2 = (z^2-\xi_1^2)(z^2-\xi_2^2)(z^2-\xi_3^2),
\end{gather*}
where we assume $0<\xi_1<\xi_2<\xi_3$, which  is easily related to
the form given by Jacobi for the case of reduction:
\begin{gather*}
w^2 = z(1-z)(z-a)(z-b)(z-ab),
\end{gather*}
our calculation shows that the associated theta function can be
put in the form
\begin{gather*}
\nonumber\Theta\left(\left[
\begin{matrix}
v_1\\ v_2
\end{matrix}\right],\tau\right)
= \frac12\theta_3\biggl(\frac12 v_1,\frac12 \tau_1\biggr)
\theta_3\biggl(\frac12 v_1-v_2,\frac12\tau_2\biggr) +
\frac12\theta_4\biggl(\frac12v_1, \frac12 \tau_1\biggr)
\theta_4\biggl(\frac12 v_1-v_2,\frac12 \tau_2\biggr),
\end{gather*}
where the $\theta_i$ are the standard $g=1$ Jacobi theta
functions, $\tau_1$ is the $\tau$ associated with the elliptic
curve
\begin{gather*}
w_1^2 = (x-\xi_1^2)(x-\xi_2^2)(x-\xi_3^2)
\end{gather*}
and $\tau_2$ is the $\tau$  associated with the elliptic curve
\begin{gather*}
w_2^2 = x(x-\xi_1^2)(x-\xi_2^2)(x-\xi_3^2).
\end{gather*}
These two $ \tau$'s can be expressed explicitly in terms of
elliptic integrals of the f\/irst kind $K$.

We adopt the homology basis shown in Fig. 1.
\begin{figure}[ht]
\begin{center}
\includegraphics[width=7cm]{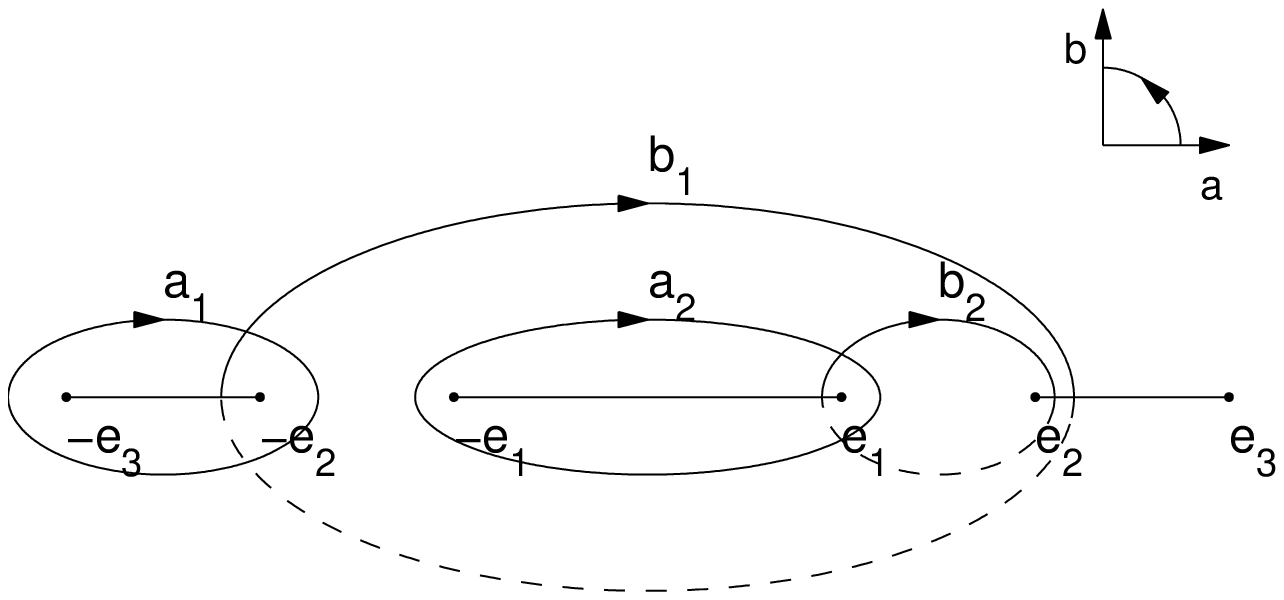}
\caption{}
\end{center}
\end{figure}

We f\/ind that
\begin{gather*}
2 \Omega_{1,1}=\int_{a_1}\frac{dz}{w}=-\omega_2, \qquad
2 \Omega_{1,2}=\int_{a_2}\frac{dz}{w}=2\omega_2,\\
2 \Omega_{2,1}=\int_{a_1}\frac{zdz}{w}= \omega_1, \qquad
2 \Omega_{2,2}=\int_{a_2}\frac{zdz}{w}=0,\\
2 \Omega_{1,1}'=\int_{b_1}\frac{dz}{w}=0, \qquad
2 \Omega_{1,2}'=\int_{b_2}\frac{dz}{w}=\omega_2',\\
2 \Omega_{2,1}'=\int_{b_1}\frac{zdz}{w}= 2\omega_1', \qquad
2 \Omega_{2,2}=\int_{b_2}\frac{zdz}{w}= \omega_1',
\end{gather*}
where the $\omega_i$ are calculated from the the homology basis
shown in Figs. 2 and 3,
\begin{gather*}
\omega_1 = \int_{\xi_2^2}^{\xi_3^2}\frac{dx}{w_1}, \qquad
\omega_1' = \int_{\xi_1^2}^{\xi_2^2}\frac{dx}{w_1},\qquad
\omega_2 = \int_{\xi_2^2}^{\xi_3^2}\frac{dx}{w_2},\qquad
\omega_2' = \int_{\xi_1^2}^{\xi_2^2}\frac{dx}{w_2}.
\end{gather*}

\begin{figure}[h]
\begin{minipage}{75mm}
\centerline{\includegraphics[width=7cm]{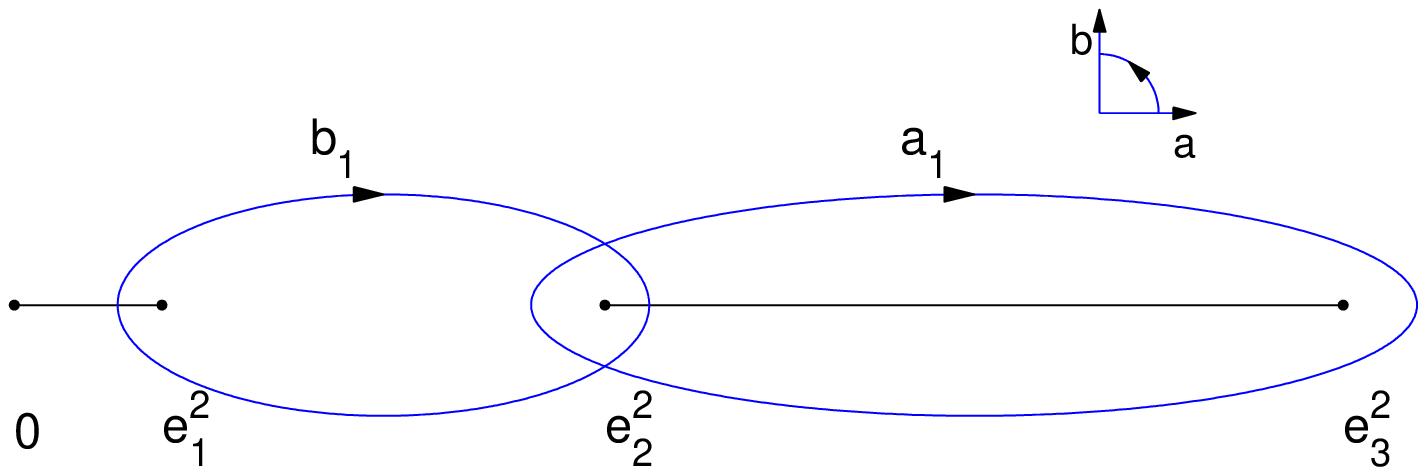}}
\vspace{-2mm} \caption{}
\end{minipage}\hfil\quad\hfil
\begin{minipage}{75mm}
\centerline{\includegraphics[width=7cm]{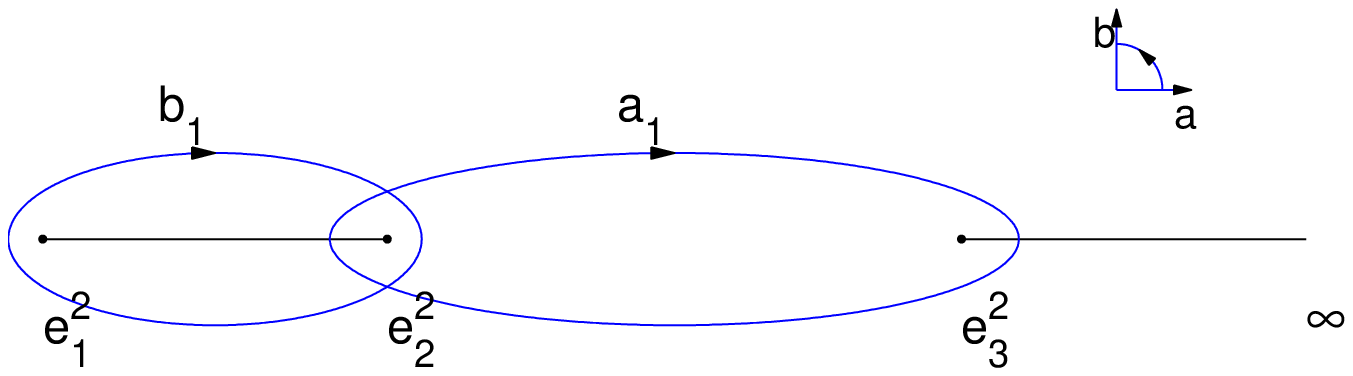}}
\vspace{-2mm} \caption{}
\end{minipage}
\end{figure}

\subsection*{Proof}
Using the change of variable $z^2=x$, $2zdz=dx$, we have
\begin{gather*}
2\Omega_{1,1}=\int_{a_1}\frac{dz}{w}=\int_{-e_3}^{-e_2}\frac{2dz}{w}
=
\int_{e_3^2}^{e_2^2}\frac{dx}{\sqrt{x(x-\xi_1^2)(x-\xi_2^2)(x-\xi_3^2)}}
\\ \phantom{2\Omega_{1,1}}
{}=-\int_{e_2^2}^{e_3^2}\frac{dx}{\sqrt{x(x-\xi_1^2)(x-\xi_2^2)(x-\xi_3^2)}}
= - \omega_2,
\\
2\Omega_{1,2}=\int_{a_2}\frac{dz}{w}=
\int_{-e_1}^0\frac{2dz}{w}+\int^{e_1}_0\frac{2dz}{w}=2\int_0^{e_1^2}
\frac{dx}{\sqrt{x(x-\xi_1^2)(x-\xi_2^2)(x-\xi_3^2)}}=
\\ \phantom{2\Omega_{1,2}}
{}=2\int_{e_2^2}^{e_3^2}\frac{dx}{\sqrt{x(x-\xi_1^2)(x-\xi_2^2)(x-\xi_3^2)}}=
2\omega_2,
\\
2\Omega_{2,1}=\int_{a_1}\frac{zdz}{w}=\int_{-e_3}^{-e_2}\frac{2zdz}{w}
=
 \int_{e_3^2}^{e_2^2}\frac{dx}{\sqrt{(x-\xi_1^2)(x-\xi_2^2)(x-\xi_3^2)}}
\\ \phantom{2\Omega_{2,1}}
{} =
-\int_{e_2^2}^{e_3^2}\frac{dx}{\sqrt{(x-\xi_1^2)(x-\xi_2^2)(x-\xi_3^2)}}
 = - \omega_1,
\\
2\Omega_{2,2}=\int_{a_2}\frac{zdz}{w} =
\int_{-\xi}^0\frac{2zdz}{w} +
\int^{\xi}_0\frac{2zdz}{w} = 0 \qquad \text{by symmetry
under}\qquad  z\rightarrow -z,
\end{gather*}
and
\begin{gather*}
2\Omega_{1,1}'=\int_{b_1}\frac{dz}{w}= 0\qquad \text{(by
symmetry)},
\\
2\Omega_{1,2}'=\int_{b_2}\frac{dz}{w}=\int_{e_1}^{e_2}\frac{2dz}{w}
=
\int_{e_1^2}^{e_2^2}\frac{dx}{\sqrt{x(x-\xi_1^2)(x-\xi_2^2)(x-\xi_3^2)}}=\omega_2,
\\
2\Omega_{2,1}'= \int_{b_1}\frac{zdz}{w}=
2\int_{e_1}^{e_2}\frac{2zdz}{w}= 2\omega_1\qquad \text{(as
for $\Omega_{2,2}'$)},
\\
2\Omega_{2,2}'=\int_{b_2}\frac{zdz}{w} =
\int_{e_1}^{e_2}\frac{2zdz}{w} =
\int_{e_1^2}^{e_2^2}\frac{dx}{\sqrt{(x-\xi_1^2)(x-\xi_2^2)(x-\xi_3^2)}}=\omega_1.
\end{gather*}
The $\omega$ can be calculated by standard integrals
\begin{gather*}
\omega_1=\int_{e_2^2}^{e_3^2}\frac{dx}
{\sqrt{(x-\xi_1^2)(x-\xi_2^2)(x-\xi_3^2)}}=
\frac{2i}{\sqrt{\xi_3^2-\xi_1^2}}\,
K\left(\sqrt{\frac{\xi_3^2-\xi_2^2}{\xi_3^2-\xi_1^2}}\right),
\\
\omega_1'=\int_{e_1^2}^{e_2^2}\frac{dx}
{\sqrt{(x-\xi_1^2)(x-\xi_2^2)(x-\xi_3^2)}}=
\frac2{\sqrt{\xi_3^2-\xi_1^2}}\,
K\left(\sqrt{\frac{\xi_2^2-\xi_1^2}{\xi_3^2-\xi_1^2}}\right),
\\
\omega_2=\int_{e_2^2}^{e_3^2}\frac{dx}
{\sqrt{x(x-\xi_1^2)(x-\xi_2^2)(x-\xi_3^2)}}=
\frac{2i}{\sqrt{(\xi_3^2-\xi_1^2)\xi_2^2}}\,
K\left(\sqrt{\frac{(\xi_3^2-\xi_2^2)\xi_1^2}
{(\xi_3^2-\xi_1^2)\xi_2^2}}\right),
\\
\omega_2'=\int_{e_1^2}^{e_2^2}\frac{dx}
{\sqrt{x(x-\xi_1^2)(x-\xi_2^2)(x-\xi_3^2)}}=
\frac2{\sqrt{(\xi_3^2-\xi_1^2)\xi_2^2}}\,
K\left(\sqrt{\frac{(\xi_2^2-\xi_1^2)\xi_3^2}
{(\xi_3^2-\xi_1^2)\xi_2^2}}\right)
\end{gather*}
so the period matrices we obtain can be written in the form
\begin{gather*}
\Omega= \left[\begin{matrix}
    -\dfrac12 \omega_2  & \omega_2\\
    \omega_1 & 0
\end{matrix}\right], \qquad
\Omega'= \left[\begin{matrix}
    0 & \dfrac12 \omega_2'\\[2ex]
    \omega_1' & \dfrac12 \omega_1'
\end{matrix}\right],
\end{gather*}
so
\begin{gather*}
\tau =  \left[\begin{matrix} 2\dfrac{\omega_1}{\omega_1'} &
\dfrac{\omega_1}{\omega_1'}
    \\[2ex]
\dfrac{\omega_1}{\omega'} & \dfrac12
\dfrac{\omega_2}{\omega_2'}+\dfrac12 \dfrac{\omega_1}{\omega_1'}
\end{matrix}\right] = \left[\begin{matrix}
    2\tau_1 & \tau_1\\[2ex]
    \tau_1 & \dfrac12 \tau_2+\dfrac12 \tau_1
\end{matrix}\right].
\end{gather*}
Following Martens, if we def\/ine
\begin{gather*}
m=\left[\begin{matrix} 0 & 0 & 2 & 1\\
1 & 0 & 0 & 0\end{matrix}\right]
\end{gather*}
then
\begin{gather*}
\left[\begin{matrix}\omega_1 & \omega_1'\end{matrix}\right]\cdot m
= 2 \left[\begin{matrix}0 & 1\end{matrix}\right] \cdot
\left[\begin{matrix}\Omega & \Omega' \end{matrix}\right].
\end{gather*}
Now def\/ine
\begin{gather*}
J=\left[\begin{matrix}0_2 & I_2
\\
-I_2 & 0_2\end{matrix}\right], \qquad T = \left[\begin{matrix} 0 &
0 &  2 &  1
\\
1 &  -2 &  0 &  2
\\
0 &  -1 &  0 &  1
\\
0 &  0 &  1 &  0 \end{matrix}\right],
\end{gather*}
where $T$ is symplectic, $T\cdot J\cdot T^t=J$, and
\begin{gather*}
m\cdot T^{-1}=\left[\begin{matrix} 1 & 0 & 0 & 0
\\
0 & 1 & -2 & 0\end{matrix}\right]
\end{gather*}
is in (Martens) standard form, and
\begin{gather*}
m\cdot J\cdot m^t=\left[\begin{matrix}0 & -2
\\
2 & 0 \end{matrix}\right]
\end{gather*}
shows that the Hopf number is 2.

The transformed matrix $\tilde \tau$ is calculated by f\/irst
calculating the $2 \times4$ matrix
\begin{gather*}
\left[\begin{matrix}A& B
  \end{matrix}\right]=\left[\begin{matrix}I_2& \tau
  \end{matrix}\right]\cdot (J\cdot T)^{-1}
\end{gather*}
then
\begin{gather*}
\tilde \tau=A^{-1}B=\left[\begin{matrix} \dfrac12 \tau_1 &
\dfrac12
\\[2ex]
\dfrac12 & -\dfrac{1}{2(2+\tau_2)}\end{matrix}\right],
\end{gather*}
where $\tau_i=\omega_i'/\omega_i$.

 For this matrix, we use Martens' transformation formula,
generalized to non-zero characte\-ristics:
\begin{gather*}
  \Theta \left[\begin{matrix}
  0 &0
  \\[1ex]
  {\dfrac12} &0\end{matrix} \right]
\left(\left[\begin{matrix} -\dfrac12v_1
  \\[1ex]
-\dfrac {v_1-2 v_2}{2(2+\tau_2)}\end{matrix}\right], \tilde \tau
\right) =\Theta\left[\begin{matrix}0
\\[1ex]
{\dfrac12}\end{matrix}\right]\biggl(-\frac{v_1}2,\frac{\tau_1}2\biggr)
\Theta\left[\begin{matrix}0
\\[1ex]
0\end{matrix}\right]\biggl(\frac{v_1-2v_2}{2+\tau2},
    -\frac{2}{2+\tau_2}\biggr)
\\ \qquad
{}+\Theta\left[\begin{matrix} 0
\\[1ex]
0\end{matrix}\right]\biggl(-\frac{v_1}2,\frac{\tau_1}2\biggr)
\Theta\left[\begin{matrix}\dfrac12
\\[1ex]
0\end{matrix}\right]\biggl(\frac{v_1-2v_2}{2+\tau_2},
-\frac{2}{2+\tau_2}\biggr)
\\ \qquad
{} = \dfrac12\theta_3\biggl(\dfrac12 v_1,\dfrac12 \tau_1\biggr)
\theta_3\biggl(\dfrac12 v_1-v_2,\dfrac12\tau_2\biggr) +
\dfrac12\theta_4\biggl(\dfrac12v_1, \dfrac12 \tau_1\biggr)
\theta_4\biggl(\dfrac12 v_1-v_2,\dfrac12 \tau_2\biggr),
\end{gather*}
in Jacobi notation, concluding the proof.

\section{Genus three}\label{sec4}

In \cite{[EEP]}, we calculated, using Halphen's ansatz as
described in Section \ref{sec1} (cf.~\cite{[U]} as well), the
Halphen spectral curve, trigonal of genus 3, for the operator:
$L=\partial^3 -15\wp\partial - ({15}/{2})\wp' =0$:
\begin{gather*}
w^3=\biggl(z^2+\dfrac{25}{4}g_3\biggr)\biggl(z^2-\dfrac{135}{4}g_3\biggr).
\end{gather*}
We showed that this curve not only admits reduction, but also has
Jacobian isogenous to the product of three (isomorphic) elliptic
curves. Here we consider the covers in the slightly more general
genus 3 curve $w^3=(z^2-\lambda_1^2)(z^2+\lambda_2^2)$,
$\lambda_1,\lambda_2 \in {\Bbb R}$, to curves of genus 1 and 2
and derive the corresponding $\tau$ matrices and reduction theory.
On occasion we indicate results for the special case
$\lambda_1^2=(135/3) g_3$, $\lambda_2^2=(25/4) g_3$:
\begin{gather*}
\tilde{\cal{C}}_3: \ w^3=\biggl(z^2+\dfrac{25}{4}
g_3\biggr)\biggl(z^2-\dfrac{135}{4} g_3\biggr).
\end{gather*}
This genus 3 curve covers the equianharmonic elliptic curve
$(\wp')^2=4\wp^3-g_3$ in three dif\/ferent ways and all entries to
the period matrices are expressible in terms of the
modulus of this curve. For the specif\/ic curve
$\tilde{\cal{C}}_3$, we have three covers, all covering
$(\wp')^2=4\wp^3-g_3$, as follows.

 The cover $\pi_1$ is given by
\begin{gather*}
  \wp =\frac{w^2}{25}\frac{(16z^2+8100 g_3)}{(4z^2-135g_3)^2},\qquad
  \wp' = \frac2{125}\frac{z(16z^4-19000z^2 g_3-759375 g_3^2)}{(4z^2-135g_3)^2}
\end{gather*}
with holomorphic dif\/ferentials given by
\begin{gather*}
\frac{dz}{w}=\frac{3}{5} \frac{d\wp}{\wp'}.
\end{gather*}

The cover $\pi_2$ is given by
\begin{gather*}
  \wp =\frac{\sqrt[3]{5}w}{20\sqrt[3]{-g_3}},\qquad
  \wp' = \frac{4z^2-55 g_3}{80 \sqrt[3]{-g_3}}
\end{gather*}
with holomorphic dif\/ferentials given by
\begin{gather*}
\dfrac{zdz}{w^2}=\dfrac{3d\wp}{4\root 3\of{5}\root
6\of{-g_3}\wp'}.
\end{gather*}

The cover $\pi_3$ is given by
\begin{gather*}
\wp = -\frac {1}{4800}\frac {w\sqrt [3]{5}
(64{z}^{6}-80\,g_{3}{z}^{4}-5300g_3^{2}{z}^{2}-30375g_3^{3})} {
(-g_{3})^{4/3}{z}^{2}(4\,{z}^{2}+25\,g_{3})},
\\
\wp' = \frac {-i p_{10}}{1152000 (4{z}^{2}+25g_{3})
{z}^{3}{g_{3}}^{2}},
\end{gather*}
where
\begin{gather*}
p_{10}=\big(-61509375\,{g_{3}}^{5}-
15187500{z}^{2}{g_{3}}^{4}-700000{z}^{4}{g_{3}}^{3}-240000{z}^{6}{g_{3}}^{2}
\\ \phantom{p_{10}=}
{}-19200{z}^{8}g_{3}+1024{z}^{10}\big) \sqrt{15}
\end{gather*}
with holomorphic dif\/ferentials given by
\begin{gather*}
\frac{dz}{w^2}=\frac{i{d\wp}\sqrt {15}\,{5}^{2/3}}{250\wp
(-g_{3})^{2/3}}.
\end{gather*}

The general curve ${\cal C}_3$ is a covering of the equianharmonic
 elliptic curve
$\cal{C}_1$ given by the equation
\begin{gather*}
\nu^2=4\mu^3+(\lambda_1^2+\lambda_2^2)^2.
\end{gather*}
The cover $\pi$ is given by
\begin{gather*}
\mu=w, \qquad \nu = 2z^2+\lambda_2^2- \lambda_1^2,
\end{gather*}
with holomorphic dif\/ferentials given by
\begin{gather*}
\dfrac{2 z dz}{3w^2}=\dfrac{d\mu}{\nu}.
\end{gather*}

\subsection{Reduction}
In this section we follow a similar approach to that of Matsumoto
\cite{[Mat]} in a genus 4 problem, who developed earlier results
of Shiga \cite{[S]} and Picard \cite{[P]}.  Write the equation of
${\cal C}$ in the form
\begin{gather*}
w^3=\prod_{k=1}^4(z-\lambda_k)=(z-\lambda_1)(z+\lambda_1)
(z-i\lambda_2)(z+i\lambda_2).
\end{gather*}
We f\/ix the following lexicographical ordering of independent
canonical holomorphic dif\/ferentials of ${\cal C}_3$, ${d}u_1=
{{d} z/w}$, ${d}u_2= {{d} z/w^2}$, ${d}u_3= {z{d} z/w^2},$ and
will def\/ine the period matrix based on the branch cuts given in
Figs.~4 and 5.

\begin{figure}[h]
\begin{minipage}{75mm}
\centerline{\includegraphics[width=7cm]{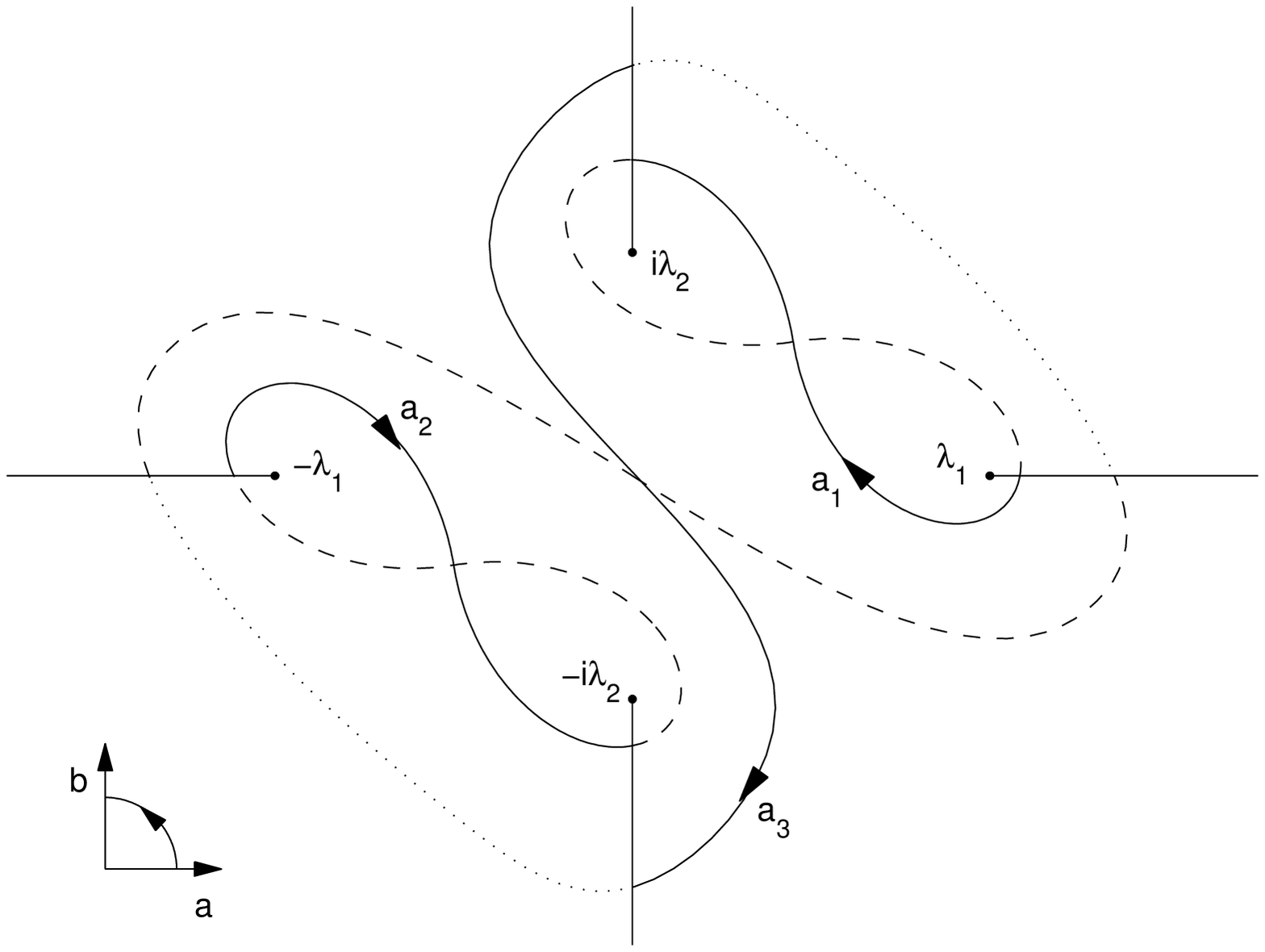}}
\vspace{-2mm} \caption{}
\end{minipage}\hfil\quad\hfil
\begin{minipage}{75mm}
\centerline{\includegraphics[width=7cm]{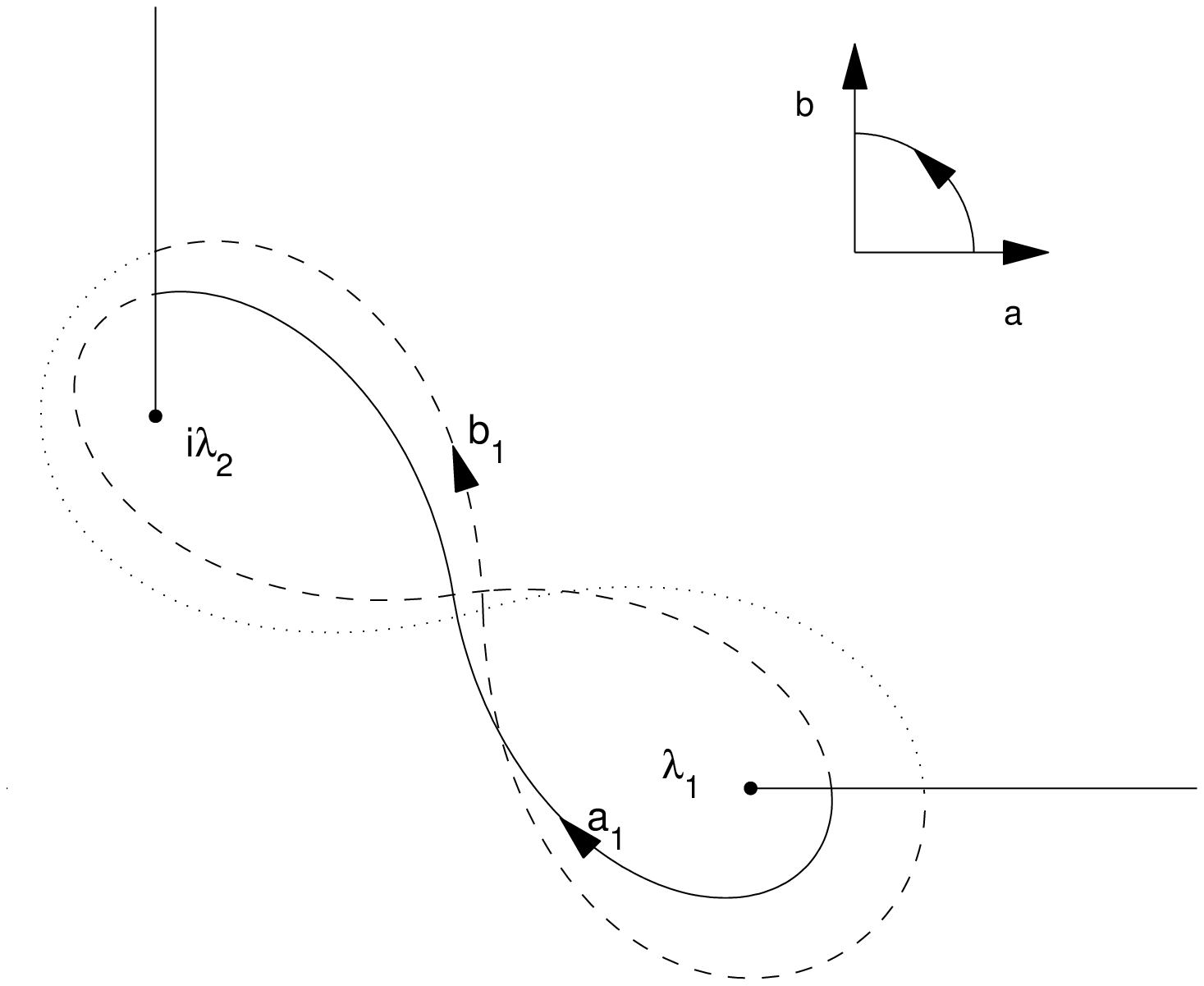}}
\vspace{-2mm} \caption{}
\end{minipage}
\end{figure}

\subsection{Riemann period matrix}

We introduce the following vector notation:
\begin{gather*}
 {\bf x}=(x_1,x_2,x_3)^T =\left(\oint_{{a}_1}{d}u_1,\ldots,\oint_{{a}_3}{d}u_1 \right)^T,
\\
  {\bf b}=(b_1,b_2,b_3)^T =\left(\oint_{{a}_1}{d}u_2,\ldots,\oint_{{a}_3}{d}u_2 \right)^T,
\\
{\bf c}=(c_1,c_2,c_3)^T
=\left(\oint_{{a}_1}{d}u_3,\ldots,\oint_{{a}_3}{d}u_3 \right)^T.
\end{gather*}

Then the matrices of ${a}$ and ${b}$-periods read
\begin{gather}
\mathcal{A}=\Biggl(\oint_{\mathfrak{a}_k}\mathrm{d}u_i\Biggr)_{i,k=1,\ldots,3}
=(\boldsymbol{x},\boldsymbol{b},\boldsymbol{c}),
\\
 \mathcal{B}=\Biggl(\oint_{\mathfrak{b}_k}\mathrm{d}u_i\Biggr)_{i,k=1,\ldots,3}
= (\rho H\boldsymbol{x},\rho^2 H\boldsymbol{b},\rho^2
H\boldsymbol{c}),
\end{gather}
where $H={\rm diag}(1,1,-1)$. The Riemann bilinear relation says
\begin{gather*}
{\bf x}^TH{\bf b}={\bf x}^TH{\bf c}=0.
\end{gather*}

The Riemann period matrix $\tau={\cal{A}}{\cal{B}}^{-1}$ belongs
to the Siegel upper half-space ${\Bbb H}^3$.

By using the symmetries of the problem we can express all the
${\bf x}$ period integrals in terms of the two integrals (see Fig.
6)
\begin{gather*}
I  =\int_0^{\lambda_1}\dfrac{dz}{w}, \qquad J
=\int_0^{i\lambda_2}\dfrac{dz}{w}.
\end{gather*}
\begin{figure}[ht]
\begin{center}
\includegraphics[width=7cm]{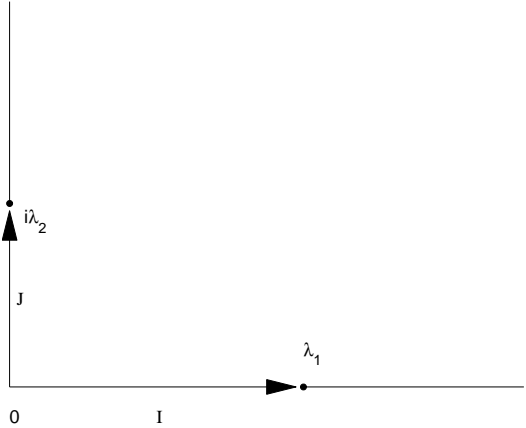}
\caption{Paths of $I$ and $J$ integrals on sheet 1.  To get the
corresponding integrals on the other sheets we multiply by $\rho$
and $\rho^2$ respectively.  The $a$ and $b$ contours can be
constructed by pasting together various combinations of these
integrals and their ref\/lections in the real and imaginary axes.}
\end{center}
\end{figure}

 We have
\begin{gather*}
     x_1  =(\rho^2-1)(I-J),\qquad
     x_2  = -x_1,\qquad
x_3  =-2J+\rho(J-I)+2\rho^2 I+\rho(J-I),
\end{gather*}
where $\rho=\exp\{2i\pi/3\}$.  In the case when
$\lambda_2^2/\lambda_1^2=5/27$, the integrals simplify further,
since $I=-J(1+2\rho)/3$. In this case (which we assume in all that
follows) we have
\begin{gather*}
 x_1  =2J(\rho+1), \qquad x_2=-x_1, \qquad x_3=2J(\rho-2).
\end{gather*}
So we have
\begin{gather*}
\left[ (\Omega')^T  \Omega^T  \right]=
\left[ \begin{matrix} -2\,J&2J&2J (3\rho+1) &2J \rho_1 &-2J \rho_1
&2J (\rho-2)
\\[.5ex]
- \rho_1 b_ {{1}}&- \rho_1 b_{2}& \rho_1 b_{3}&b_{1}&b_{2}&b_{3}
\\[.5ex]
- (\rho+1) c_{1} &-(\rho+1) c_{2}& \rho_1
c_{3}&c_{1}&c_{2}&c_{3}\end{matrix} \right],
\end{gather*}
where $\rho_1=\rho+1$.  We def\/ine
\begin{gather*}
\tau = ((\Omega')^T )^{-1} \Omega^T.
\end{gather*}
We f\/ind after much simplif\/ication, using the properties of
$\rho$ and the Riemann relations, that
\begin{gather*}
\tau =  \frac1{79} \left[\begin{matrix} 62\rho-13 &17\rho+13
&-5\rho+38
\\[.5ex]
17\rho+13 &62\rho-13 &5\rho-38
\\[.5ex]
-5\rho+38 &5\rho-38 &45\rho+53\end{matrix}
 \right].
\end{gather*}

This matrix $[(\Omega')^T \,\Omega^T]$ satisf\/ies the reduction
criteria as def\/ined by Martens \cite{[Ma]}, since if we def\/ine
\begin{gather*}
  \Pi =\left[ \begin{matrix}2J & 2J\rho \end{matrix} \right],\qquad
   H =\left[ \begin{matrix}1 &0&0 \end{matrix} \right],
\end{gather*}
then
\begin{gather*}
H\cdot \left[ (\Omega')^T \ \Omega^T \right]=\Pi \cdot M,
\end{gather*}
where
\begin{gather*}
M= \left[ \begin{matrix} -1&1&1&1&-1&-2
\\[.5ex]
0&0&3&1&-1&1\end{matrix} \right].
\end{gather*}
We can transform $M$ to standard form using the symplectic matrix
\begin{gather*}
S =  \left[ \begin{matrix} 1&3&-1&0&-6&2
\\[.5ex]
0&-1&-1&0&1&-1
\\[.5ex]
0&1&0&-3&-1&0
\\[.5ex]
0&0&0&1&0&0
\\[.5ex]
0&0&0&-1&0&-1
\\[.5ex]
0&-2&0&2&3&-1
\end{matrix} \right],
\end{gather*}
since
\begin{gather*}
M.S=\left[ \begin {array}{cccccc}
-1&0&0&0&0&0\\\noalign{\medskip}0&1&0&- 5&0&0\end {array} \right]
.
\end{gather*}

Following Martens, the corresponding transformed $\tau$ is
\begin{gather*}
\tilde \tau = \tau \cdot S \cdot J^T= \left[\begin{matrix}
\dfrac15+\dfrac15\,\rho&\dfrac15&0
\\[2ex]
\dfrac15 & \dfrac12+\dfrac15\,\rho&\dfrac12\,\rho
\\[2ex]
0& \dfrac12\,\rho&\dfrac32\,\rho+\dfrac12\end{matrix}
 \right].
\end{gather*}
Expanding a theta function def\/ined with this $\tilde \tau$,
again following Martens, we will get a sum of 5~products of $g=1$
theta functions with $g=2$ theta functions.  The $g=1$ theta
functions will have tau value $\tau=(1/5)(1+\rho)$ and the $g=2$
theta functions will have a tau of
\begin{gather*}
\tau =\left[ \begin{matrix} 5 & 0
\\[2ex]
0 & 1\end{matrix}\right] \left[\begin{matrix}
\dfrac12+\dfrac15\,\rho&\dfrac12\,\rho
\\[2ex]
\dfrac12\,\rho&\dfrac32\,\rho+\dfrac12
\end{matrix} \right]\left[\begin{matrix}
5 & 0
\\[2ex]
0 & 1\end{matrix}\right]= \left[\begin{matrix} 5\,\rho+{\dfrac
{25}{2}}&\dfrac52\,\rho
\\[2ex]
\dfrac52\,\rho&\dfrac32\,\rho+\dfrac12
\end{matrix} \right].
\end{gather*}
Again following Martens, we can reduce this $2\times 2$ matrix to
standard form to get eventually
\begin{gather*}
\tilde\tilde \tau=\left[ \begin {array}{cc} \dfrac{11}{20}+
\dfrac1{20}\,\rho & -\dfrac14
\\
\noalign{\medskip}-\dfrac14 & \dfrac32+\dfrac14\,\rho\end {array}
\right] .
\end{gather*}
Expanding each of the transformed genus 2 theta functions with
this theta will give a product of genus 1 theta functions with
$\tau=(1/20)(11+\rho)$ and genus 1 theta functions with
$\tau=16({3/2}+(1/4)\,\rho)$. So f\/inally we have $5\cdot 4=20$
terms, each containing a product of three genus~1 theta functions
(with fractional characteristics).

\section{Conclusions} 

We contributed to the theory of spectral curves of ODOs with
elliptic coef\/f\/icients routine algorithms to calculate:
\begin{itemize}
\itemsep=0pt \item The algebraic equation of the curve (always, in
principle);

\item The period matrix (only if the periods can be chosen
suitably and there is an explicit solution to the action
equations);

\item A reduction method for the theta function.
\end{itemize}

What remains to be calculated (Challenge IV) is the dependence of
the coef\/f\/icients on the time parameters. This is more
dif\/f\/icult because it involves expanding an entire basis of
dif\/ferentials of the f\/irst kind.  In \cite{[EEP]}, we were
able to f\/ind the time dependence by implementing Jacobi
inversion, thanks to the Hamiltonian-system theory which describes
the evolution of the poles of the KP solution \cite{[Kr]}.


\pdfbookmark[1]{References}{ref}
\LastPageEnding

\end{document}